\documentclass[11pt]{article}
\usepackage{epsfig,rotating}
\usepackage{color}
\usepackage{amsmath}
\protect{\markright{Phillips}}

\begin{document}
\baselineskip 24pt
\title{When is Better Best? A multiobjective perspective}
\author{Mark H Phillips, Ph.D. \\
Clay Holdsworth, Ph.D. \\
\\
Department of Radiation Oncology, \\
University of Washington Medical Center, \\
Box 356043, Seattle, WA 98195, USA  \\
}


\maketitle

\normalsize
\newpage
\begin{abstract}
\noindent \underline{Purpose:} To identify the most informative
methods for reporting results of treatment planning comparisons.   \\ 
\noindent\underline{Methods~and~Materials:} Seven papers from the past
year of {\it International Journal of Radiation Oncology Biology
  Physics} reported on comparisons of treatment plans for IMRT and
IMAT.  The papers were reviewed to identify methods of comparisons.
Decision theoretical concepts were used to evaluate the study methods
and highlight those that provide the most information.  \\
\noindent\underline{Results:} None of the studies examined the
correlation between objectives.  Statistical comparisons provided some
information but not enough to make provide support for a robust
decision analysis.  \\
\noindent\underline{Conclusion:} The increased use of treatment
planning studies to evaluate different methods in radiation therapy
requires improved standards for designing the studies and reporting
the results.
\end{abstract}

Keywords: treatment planning, comparisons, Pareto optimal

\section{Introduction}
\label{intro}

In radiation therapy, new algorithms, delivery methods and hardware are
introduced into the marketplace and clinical practice on a regular
basis.  While a clinical trial is the benchmark for comparisons in the
field, such trials are time-consuming and expensive in manpower and
resources.  The result is that other comparison methods must
suffice.  One popular method is a comparison between the two methods
in which the bases of comparison are the
treatment plans calculated using the ``old'' and the ``new'' method.
Using a software model as the sole method of comparison for clinical
practices is relatively rare outside of radiation oncology; animal or
human trials (prospective or retrospective) are the norm in most other
specialities.  

One of the advantages of clinical trials is that they usually produce data
that are directly connected to clinical decision-making.  Even so,
applying the results can be difficult because of the multicriteria
nature of clinical decisions.  Most often, new therapeutic methods or
diagnostic tests involve trade-offs.   A test or therapy may provide
new information or improve an outcome but it may also introduce a risk of
complications that the other tests or therapies do not have, and a
decision must be made as to whether the improvement is worth the
increased risk.  The new method may also incur costs--in manpower,
time, equipment--that can play a decisive role in the decision as to
whether to implement it.  

Whether explicitly stated or not, comparisons of different modalities
can be framed in terms of multicriteria decision making (MCDM), and therefore
it is appropriate to use the methods and concepts from this field in
analyzing such studies.  The act of comparing two or more
procedures or modalities means that one wishes to determine if one is
better than the others, and in the context of medical practice, using
this information to decide on a method of treatment.  Decision making
can be divided into four realms that characterize the type of
decisions: single or multiple criteria, and under certainty or under
uncertainty.  The relevant realm for this work is multiple criteria
under uncertainty since the decision affects 
the probabilities of the different possible outcomes, and the decision
rests on the valuation of multiple components of the problem.

Given the difficulty in conducting trials, and acknowledging the
difficulties inherent in treatment planning comparisons, what can be
expected from such studies and what are the best methods for
conducting them?  This paper provides some characteristics that would make the
results more useful and less prone to bias.  These are evaluated against a set
of papers from one year's issues of  {\it International Journal of
  Radiation Oncology Biology  Physics} which all address the
differences between intensity modulated radiation therapy (IMRT) and
intensity modulated arc therapy (IMAT).  The goal is not to come to a
conclusion about the efficacy of IMAT vs IMRT; rather
it is to investigate which comparison methods provide the most
useful information, and perhaps to stimulate discussion regarding the
most productive means for so doing.

\section{Methods}
\label{methods}

The concepts highlighted in this work have been explored in a number
of different fields, and this has resulted in a lack of
standardization with respect to terminology.  For clarity's sake, we
present some definitions.
\begin{itemize}
  \item criterion: an aspect of the decision making problem that is
    critical to the decision making process.  Also referred to in the
    literature as an {\it
      objective}, {\it attribute} or {\it goal};
  \item objective: a desired goal that can be written in mathematical
    form and is part of a mathematical optimization algorithm.  It is
    represented by an {\it objective function} and the value of this
    function for a given set of circumstances is the {\it objective
      value}.
\end{itemize}

As described above, ``objective'' is a fairly general term, and our use
is meant to highlight the difference between a criterion, which may be
difficult to write in a mathematical form, and a functional
formulation of that criterion whose mathematical character allows it
to be used in optimization and/or decision making algorithms.  In the
context of inverse planning, more detailed distinctions between types
of objectives are useful.  
\begin{itemize}
  \item inverse planning objective (IPO): an objective (in the sense given
    above) focussing on a single criterion that is used in an
    optimization algorithm to calculate beam delivery variables; 
  \item decision objective (DO): an objective (in the sense given above)
    that is used to rank plans.  It may be identical to the IPO; it
    may be the mathematical representation of a criterion; or it may
    be a surrogate for a criterion that  is difficult to
    express mathematically.
\end{itemize}

Depending on the optimization algorithm, several IPO's can be summed
or multiplied to produce a single, global function whose value is
optimized. 
There are several reasons why an inverse planning objective may not be
used as a decision objective.  Optimization algorithms are most
efficient when the objective functions are of certain forms, which may
not be close enough to the functional form of the decision making criteria.
It can also be desirable to have a limited set of DO's (as
described below), whereas the optimization may proceed better with a
larger set of objectives.  

\subsection{Comparisons in literature}

One year's issues of {\it International Journal of
  Radiation Oncology Biology  Physics} (Volumes 74 - 76) were searched
for articles on treatment planning comparisons of IMRT and
IMAT\footnote{The generic term IMAT is used for modulated arc therapy rather
  than the often-used trademarked names.}.  Seven articles were found
\cite{Tang10,Yoo10,Popescu10,Weber09,Bignardi09,Wu09,Verbakel09}.  The
comparison between these two methods was chosen because of the large
number of articles, the rapid dissemination of IMAT, and the fact that
the comparison is complex given that it involves hardware, delivery
software, planning software and products from competing vendors.  The
tumor types and sites included brain, head \& neck, lung, prostate,
breast, lymphoma, paraspinal, and lymph nodes.

Table 1 provides the relevant details extracted from these papers.
These treatment planning studies involved optimization using
inverse planning algorithms; the objectives used in these calculations
are termed {\it Inverse Planning Objectives (IPO)}.  The resulting
treatment plans were compared using another set of objectives, which
we call {\it Decision Objectives (DO)}.  

\subsection{Multicriteria decision theory}

Given the the fact that we are interested in 
multicriteria (also known as {\it multiobjective}) decisions
made under uncertainty, 
the decision making environment involves both
probabilities and values.  Values are, of course, the essential
element in deciding between alternatives and are most often
encapsulated using utility functions.  {\it Utility} is the
quantitative measure of the strength of preference for an outcome
\cite{Keeney93,Hunink01}.  Expected utility is the sum of the utilities of all
the possible outcomes  of a given action weighted by the probabilities
of their occurence.  One method for deciding is to choose that action
(among all allowed actions) that results in the maximum expected
utility.  

The method most often used to compare plans in radiation therapy is to
ignore the probabilistic nature of the outcome.  In the most common
form of inverse planning, the objective values are weighted by the
value of each objective by means of a ``weighting factor'' or
``importance factor'', and the sum of weighted objective values is the
score of the plan.  This is a powerful but limited method that quickly
finds a solution but comparisons can be difficult when the objective
functions have been changed. Current inverse planning algorithms
provide little guidance for searching for better plans and the
arbitrary objective function values insure that the weighting factors
cannot really serve as  utilities.


As described earlier, nearly all comparisons of interest are
multicriteria problems.  In many cases, the criteria are not
commensurate, but even when they are our sense of determining the
``best'' outcome means that they must be synthesized into a single
metric.  Expected utility and the weighting factor method, as
described above, are such 
metrics but it can be difficult to apply them when determining
preferences for the trade-offs that are inherent in the problem.  In
the absence of a utility or value function, goal programming is a
useful approach.  This approach seeks to achieve each of the desired
goals and only requires that the decision maker can set appropriate
goals for each criterion.  Outcomes can be ranked by the proportion of
criteria (goals) that are met, assuming that the goals are not too
easy to achieve.

Finally, an aspect of decision making that is critical for the
problems of interest is the situation in which
the criteria are linked or correlated, especially when a good value
for one results in a poor value for another.  That means that
it makes little sense to compare single criteria against each other
since the connections with the other criteria are lost.  In the case of
planning comparisons, it is quite possible to have a method provide
better values for one criterion while having poor values for
other, correlated criteria.  

One measure
that does take the correlation into account is {\it domination}.  If
the outcomes for $k$ objectives,
$\vec{x}~=~\{x_1,x_2,\mathellipsis,x_k\}$ are compared for two
  different situations, {\bf x'} and {\bf x''}, then {\bf x'} is said
  to dominate {\bf x''} if:
\begin{equation}
\begin{split}
  & x_i' \le x_i'' ~\forall ~i~ \in~ k \\
& \text{as long as}~ x_i' < x_i''~ \text{for at least one value of } i \\
&  \text{where it is assumed that a lesser objective value is preferred
   over a greater.}
\end{split}
\end{equation}
 
A solution that dominates another must be considered better, within the
preference space defined by the chosen objectives, since there would
be no reason to prefer the solution with all higher (less desirable)
values.  This is 
similar to, but not identical with, Pareto optimal solutions or the
Pareto front since  Pareto optimality means that there exist no
better solutions.  Many optimization algorithms produce Pareto optimal
plans given the IPO's used in the optimization, although it is likely
that they are not Pareto optimal in DO space.  The mismatch between
IPO space and DO space is one of the difficulties in performing
inverse planning.

 The advantage of the domination comparison
is that no values or rankings between the multiple objectives need be
assigned.   Two methods can be compared by calculating the number of
outcomes, i.e. plans, produced by one method that dominate the
outcomes (plans) from the second method, 
and vice versa.  If one method results in a greater number of dominant
plans, then that method has a clear advantage.   An important point to
note is that the number of objectives that is used to characterize a
plan greatly influences the number of dominations.  With large numbers
of objectives, it is highly likely that while one plan may be superior
in many objectives compared to another plan, there will be one or two
objectives in which it is not, thereby preventing domination.   

Strict comparisons of dosimetric objectives may be modified
to account for clinical significance in the sense that a
difference of a percent or less in the dose to some critical structure
may not mean that one plan is superior in any clinically meaningful
way.  The domination comparison can be 
modified to include an $\epsilon$ of such a magnitude that when the
difference between two objective values is less than $\epsilon$, the
values are considered clinically equal.  The size of $\epsilon$
depends on the objective of interest; the recent publication of
normal tissue responses \cite{Quantec10} provides a useful  compendium of
uncertainties in complication probabilities for a range of tissues.    

While domination comparison is the most direct way to determine
which method is better, it is not always easy or appropriate.
For one, statistical measures of outcome differences require a large
number of plans which may not be feasible.  They also operate best
when a small number of objectives can be selected for comparison.  If
the method being compared is exploring clinically new territory, it
may be difficult to decide on a small number of the most critical
criteria.  It may also be the case that the 
investigator may not wish to impose a set of values when it is
acknowledged that different practitioners may have different priorities.
In these cases, providing probabilities for the different outcomes
provides the reader with a useful set of numbers that they can then
apply as they see fit.    The correlation between opposing objectives
needs to be part of the calculation.  For 
example, a method may produce plans that provide
better tumor coverage 50\% of the time and better OAR sparing 50\% of
the time.  However, the chance that it produces better tumor coverage
and OAR sparing in a given plan may be anywhere from 0 to 50\%.  
Therefore, the appropriate probabilities would include at least two
relevant, conflicting conditions, e.g. the probability that a plan
achieves a certain tumor coverage and maintains an OAR dose below a
certain critical level or the probability that the plan maintains a
high minimum tumor dose and good tumor dose homogeneity.  

If probability distributions are obtained, the methods can be compared
by examining the integral probability distribution.  If the
distribution from one method does not cross that of the comparison
method, then one method ``stochastically dominates'' the other and can
be considered superior.
 
In these comparisons, there are two levels of pairing of results.  At
one level, results for each patient case must be compared against each
other in order to avoid artifacts dependent on the particular
cases. The second level is that objectives are often paired with 
competing objective(s) in order to capture how each method deals with
the inherent trade-offs.  In  summary, the most appropriate methods
for comparing different 
modalities are: \\
\vspace{-12pt}
\begin{itemize}
\item Probabilities
  \begin{itemize}
    \item determine conflicting DO's 
    \item find joint probabilities for each set of DO's  
  \end{itemize}
\item Expected Utility 
  \begin{itemize}
    \item find probabilities that each method achieves each DO
    \item determine utilities 
    \item calculate which method yields highest expected utilities
  \end{itemize}
\item Domination Comparison 
  \begin{itemize}
    \item case by case comparison of all DO values
    \item count proportion of cases in which DO values of one method are
      as good as or better than those of second method
  \end{itemize}
\item Goal Programming
  \begin{itemize}
    \item set goals for each DO value
    \item count how many goals are met by each plan 
  \end{itemize}
\end{itemize}


\section{Results}
\label{results}

Table~\ref{tab:summaries} summarizes the comparisons for each of the
published studies.  The number of separate patient cases ran from 5 to
14 (in one 
paper, \cite{Tang10}, three cases were studied for four different
sites), and included multiple tumor sites.  The inverse planning
objectives were different from the decision objectives in all cases,
though there was some overlap in several studies.  The DO's included
many of the same types of objective functions used as IPO's such as
mean doses and DVH parameters $V_{x}$ and $D_{x}$.  More complex
objectives were also used as DO's, such as equivalent uniform dose
(EUD), homogeneity index (HI), conformality index (CI), and normal
tissue complication probability (NTCP).  Most of the studies also
included some measure of treatment efficiency, such as total monitor
units, total number of segments, and/or beam-on time. 

Comparisons between the modalities were made using several methods,
including individual and averaged dose-volume histograms, direct
comparison of each DO for 
each patient case, direct comparison of the means of each
DO (averaged over all patient cases) using both paired and unpaired
statistical tests, and the values of the Wilcoxon
summed rank test for each DO averaged over all patient cases.  
With respect to the statistical significance of the tests used, some
of the studies reported conflicting results with respect to 
the superiority of IMAT over IMRT or vice versa.

Reference~\cite{Tang10} concluded that IMAT obtained the best target
coverage in most cases and the lowest toxicity in some cases.  This
was the one study in which data were presented case by case, although
the competing objectives were not linked in the analysis.  From the
presented data, it was seen that IMRT dominated IMAT in two cases, and
IMAT dominated IMRT in one case. In the other nine, IMRT had better
values for some DOs and worse for others.  Of those nine, IMRT had
more superior DO values in two cases, worse in seven, and in one case
IMAT and IMRT had the same number of superior DO values.  

Reference~\cite{Yoo10} used a homogeneity index to characterize the
dose to the target volume and reported a difference of 1\% in the
means (averaged over all patient cases) and this was a statistically
significant difference.  Using the methods in Table~\ref{tab:methods},
IMRT was considered superior to IMAT for target and all OARs when one
arc was used, and when two arcs were used, IMRT was better in only 3
of 6 OARs and the targets were substantially equivalent.  Overall,
they concluded that IMAT was more efficient but IMRT was better
dosimetrically, and that one should compare the two on a case-by-case
basis before selecting one over the other.

Reference~\cite{Popescu10} concluded that there was no statistical
difference between the modalities for the targets and that IMAT was
superior for the OAR DOs. The results
for the homogeneity (HI) and conformality (CI) indices for the targets were
presented case by case.  The average HI was better for IMAT, the
average CI was better for IMRT, but in two of the cases the HI and CI
were superior for IMRT, in one case both were superior for IMAT and
the results were mixed for the other two cases.  An argument was
presented that the reduction in contralateral breast dose with IMAT
would lead to improvements in outcomes of secondary cancers.  They
concluded that IMAT was better than IMRT for these cases.

Reference~\cite{Weber09} provided means for 36 different DOs and the
value of $p$ for the Wilcoxon signed rank test.  The large number of
DOs were intended to provide  a summary of the DVH for each OAR and
the target.  A relatively small number of DOs were statistically
different (all to the benefit of IMAT), and no general conclusions
regarding the two modalities were drawn.

Reference~\cite{Bignardi09} came to the general conclusion that IMAT
produced a level of normal tissue avoidance similar to IMRT.  Explicit
evaluation criteria were described and mean dosimetric parameters were
reported.  Statistically different values were reported for some target
metrics as well as some OAR metrics, but no distinction was made with
respect to which modality performed better.  The shorter treatment
time of IMAT was an explicit objective tied to a clinical need for
improving positioning compliance.  

Reference~\cite{Wu09} reported EUD values (along with DVH metrics)
because of the inhomogeneous dose distributions.  They concluded that
IMRT and IMAT had comparable target coverage, and that IMRT resulted
in more spinal cord sparing if one arc was used, and comparable
sparing if two were used.  Only one target metric (conformity index)
showed statistical superiority, although the
direction was not stated.  The spine was the only OAR that had a
statistical difference.  A final conclusion recommended further
clinical studies to investigate the efficacy of IMAT. 

Reference~\cite{Verbakel09} concluded that single arc IMAT plans were
similar to IMRT plans except for a reduced target homogeneity.  Double
arc IMAT was judged superior in target dose and similar in OAR
sparing.  Given these results plus the reduced delivery time and fewer
monitor units, this group made the decision to replace IMRT with IMAT
for all indications.

Most of the studies made specific statements regarding the fact
that the inverse planning process or dose calculation algorithms
had the potential to introduce some bias into the outcomes.     Nearly
all of the studies also concluded that the IMAT treatment took less
time and/or fewer monitor units to deliver the treatments.


\section{Discussion}
\label{discussion}

Seven recent papers comparing IMAT to IMRT were reviewed and analyzed
to determine the extent to which they provided information consistent with
the principles of multicriteria decision theory.  In the context of
comparing different methods by means of treatment planning, 
the key principles are (1) dealing appropriately with multiple, competing
criteria, and (2) handling the uncertain nature of the outcomes. 


The papers reported results of a number of different statistical
comparisons.   One of them, the Student's t-test (unpaired),
ignores the fact that much of the variability stems from the inherent
anatomical differences between patients.  The Student's paired t-test
properly focuses on the differences in the values of the DO's, but it
is applied under the assumption that the DO values follow a normal
distribution.  Although it is natural to assume that there will be
relatively few small or large DO values, it is a difficult assumption
to prove since the distribution is based both on the anatomy of a
patient population and the inverse planning algorithm and its use. The
Wilcoxon rank test is a more general metric and is probably more
appropriate to the data in this situation. 

While the use of some of these statistics does account for the fact that
the data are pair-wise matched due to the differences in patient
cases, they ignore a critical correlation, namely the fact that many
of the objectives are linked to one another in such a way that
satisfaction of one can lead to less satisfactory values of the others.
This linkage, caused by the nature of radiation transport,
is essential to any comparison.  Only Ref.~\cite{Tang10} provided data
with which the connection between the DOs could be assessed.  
The other papers assessed differences for each DO separately by means of
the Student's t test, Student's paired t test, and the Wilcoxon
test. In the cases when both targets and OAR's yielded 
significant improvements in the DO's on average for a given method, it is
likely that for given cases, both DO values are better.  However, it
is also likely that this is not true for all cases, and may not be
true for the majority of cases.  In this respect, the Wilcoxon signed
rank test is a stronger measure than the Student's paired t test.
However, in neither case is it possible to determine any probabilities
for both values being better (or worse). 
  When the tests yielded conflicting significant
differences, e.g. better target coverage but worse OAR sparing, these
tests provide no reliable information regarding the probabilities of
relevant outcomes.  On the other hand, when there is no statistically
significant difference, it is still possible that there is an advantage
of one method over the other with respect to domination.

A crucial difference between these planning studies and other clinical
comparisons is the trade-offs between the outcomes.  In treatment planning
comparisons (and particularly those studied here), there is a
trade-off between the separate outcomes that requires that joint outcomes be
reported.  Clinical outcomes are usually perceived as being
stochastically related, and joint outcomes are not usually
reported\footnote{However, given the the large amount of knowledge yet
  to be gained about individual radiation response, one could make the
  argument that reporting the cases in which tumor and complication
  outcomes occurred jointly would be potentially useful.  }.
Therefore, the main focus of this paper has been on establishing
reporting methods that are unique to these types of studies, and
establishing standards for when better is best.  

When comparisons are made in clinical studies, the purpose of the study
is usually clearly spelled out, e.g. which method produces longer
survival or fewer complications.  In these studies, implicit values are
assigned to criteria and the method for judging one superior to the
other is based on a clear comparison of these criteria.  
 In the papers reviewed, the purposes were not nearly as explicit, only
 that the methods were to be compared.  As described above, if the
 utilities of the outcomes are not provided and the study is only
 meant to provide the bases for decisions and not the decisions
 themselves, then a statistical  distribution of linked objectives is
 needed.  If the purpose is to determine which is best, then the
 criteria need to be explicitly stated and the appropriate method
 chosen. 

A potentially important confounding factor in treatment planning
comparisons is the subjective element in the current inverse planning 
paradigm, as exemplified by several examples.  From Bignardi {\it et
  al}, ``Both IM and RA plans were 
optimized using the same objectives by the same experienced planner
aiming to respect planning strategies described above'', and from
Verbakel {\it et al}, ``All IMRT optimizations were done by
interactively adapting the objectives and their priorities.''  Inverse
planning algoritms, as currently implemented in commercial planning
systems, provide little insight or guidance in the search of better
plans \cite{Holdsworth10,Phillips10a}.  The process of searching
available plan space is very much a trial-and-error process and it is
very difficult to say whether a better plan could have been achieved.
Therefore, as several papers noted, there is considerable uncertainty
in the capabilities of the two methods.






\section{Conclusion}
\label{conclusion}

A year's worth of papers reporting comparisons between treatment plans
utilizing IMRT and those using IMAT were analyzed with respect to the
methods of comparison.  The major weakness in these studies was the
lack of coupling between the results for competing plan criteria.
The comparisons fall into the realm of multicriteria decision
making.  Applying the principles of  MCDA, it is concluded that the
results of the comparisons would be more useful if (a) the criteria
and methods  of
comparison were explicitly stated and justified, (b) the probabilities
of occurrence of the criteria were reported, and/or (c) explicit
utilities for the criteria were provided and used to rank the methods.


\begin{center}
{\sc Acknowledgments}
\end{center}

This work was supported by NIH Grant R01 CA112505.

\newpage

\begin{thebibliography}{10}

\bibitem{Tang10}
G~Tang, M~A Earl, S~Luan, C~Wang, M~M Mohiuddin, and C~X Yu.
\newblock Comparing radiation treatments using intensity modulated beams,
  multiple arcs and single arcs.
\newblock {\em Int J Radiat Oncol Biol Phys}, 76:1554--1562 (2010).

\bibitem{Yoo10}
S~Yoo, Q~J Wu, W~R Lee, and F~F Yin.
\newblock Radiotherapy treatment plans with rapidarc for prostate cancer
  inolving seminal vesicles and lymph nodes.
\newblock {\em Int J Radiat Oncol Biol Phys}, 76:935--942 (2010).

\bibitem{Popescu10}
C~C Popescu, I~A Olivotto, W~A Beckham, W~Ansbacher, S~Zavgorodni, R~Shaffer,
  E~S Wai, and K~Otto.
\newblock Volumetric modulated arc therapy improves dosimetry and reduces
  treatment time compared to conventional intensity modulated radiotherapy for
  locoregional radiotherapy of left-sided breast cancer and internal mammary
  nodes.
\newblock {\em Int J Radiat Oncol Biol Phys}, 76:287--295 (2010).

\bibitem{Weber09}
D~C Weber, N~Peguret, G~Dipasquale, and L~Cozzi.
\newblock Involved-node and involved-field volumetric modulated arc vs. fixed
  beam intensity-modulated radiotherapy for female patients with early-stage
  supra-diaphragmatic hodgkin lymphoma: A comparative planning study.
\newblock {\em Int J Radiat Oncol Biol Phys}, 75:1578--1586 (2009).

\bibitem{Bignardi09}
M~Bignardi, L~Cozzi, A~Fogliata, P~Lattuada, P~Mancosu, P~Navarria, G~Urso,
  S~Vigorito, and M~Scorsetti.
\newblock Critical appraisal of volumetric modulated arc therapy in
  stereotactic body radiation therapy for metastases to abdominal lymph nodes.
\newblock {\em Int J Radiat Oncol Biol Phys}, 75:1570--1577 (2009).

\bibitem{Wu09}
Q~J Wu, S~Yoo, J~P Kirkpatrick, D~Thongphiew, and F~F Yin.
\newblock Volumetric arc intensity–modulated therapy for spine body
  radiotherapy: Comparison with static intensity-modulated treatment.
\newblock {\em Int J Radiat Oncol Biol Phys}, 75:1596--1604 (2009).

\bibitem{Verbakel09}
W~F A~R Verbakel, J~P Cuijpers, D~Hoffmans, M~Bieker, B~J Slotman, and S~Senan.
\newblock Volumetric intensity-modulated arc therapy vs. conventional imrt in
  head-and-neck cancer: A comparative planning and dosimetric study.
\newblock {\em Int J Radiat Oncol Biol Phys}, 74:252--259 (2009).

\bibitem{Keeney93}
R~L Keeney and H~Raiffa.
\newblock {\em Decisions with Multiple Objectives}.
\newblock Cambridge University Press, Cambridge, UK (1993).

\bibitem{Hunink01}
M~Hunink and P~Glasziou.
\newblock {\em Decision making in health and medicine: integrating evidence and
  values}.
\newblock Cambridge University Press, Cambridge, UK (2001).

\bibitem{Quantec10}
L~B Marks, R~K~Ten Haken, and M~K Martel, editors.
\newblock {\em Int J Radiat Oncol Biol Phys}, volume~76, Supplement. 
  Quantitative analyses of normal tissue effects in the clinic.
\newblock ASTRO (2010).

\bibitem{Holdsworth10}
C~Holdsworth, J~Liao, M~Kim, and M~H Phillips.
\newblock A hierarchical evolutionary algorithm for multiobjective optimization
  in {IMRT}.
\newblock {\em Med Phys}, 4986-4997 (2010).

\bibitem{Phillips10a}
M~H Phillips.
\newblock How to make clinical decisions in the multicriteria framework
  (abstr).
\newblock {\em Med Phys}, 37:3403 (2010).

\end{thebibliography}

\newpage

\begin{sidewaystable}
\begin{center}
\footnotesize
\begin{tabular}{|l|l|l|l|} \hline
              & Tumor types & Inverse Planing . & Decision \\
  Reference & (\# of cases) & Objectives (I.P.O.) & Objectives (D.O.)
  \\ \hline \hline
 Tang {\it et al}\cite{Tang10} & H \& N (3) & RTOG 05-22$^1$    &
  $V_x$ (OAR$^2$), 
$V_{95\%}$ (target) \\ 
                    & brain (3) & RTOG 05-13$^1$ & \\
                    & lung (3) & RTOG 06-17$^1$  &   \\
                    & prostate (3) & RTOG 04-15$^1$  &  \\ \hline
 Yoo {\it et al}\cite{Yoo10} & prostate & $D_{100}$, $D_{max}$  (target) &
$V_{65Gy}$, $D_{mean}$ bladder  \\
                   & + seminal vesicles & $V_x$, $D_x$, $D_{max}$ for
& 
$V_{70Gy}$, $D_{mean}$  rectum  \\
                  & + pelvic lymph nodes (10) & rectum, bladder, bowel & 
$V_{45Gy}$, $D_{mean}$  bowel  \\
                   & & femoral heads, normal tissue & HI$^3$, CI$^4$ target
 \\ \hline
Popescu {\it et al}\cite{Popescu10} & left breast & $V_{95\%}$ PTV & HI
breast/chest wall   \\ 
         & + IM nodes (5) & $V_{30Gy}$ heart & $V_{42.7Gy}$ HI nodal
  PTV \\
         &                & $V_{20Gy}$ ipsilateral lung & CI each PTV   \\
   & & & CI combined PTV  \\
   & & & $D_{2\%}$ PTV  \\
   & & & $V_{xGy}$ heart, lung, contralat. breast   \\ 
   & & & mean EUD for all structures  \\ \hline 
 Weber {\it et al}\cite{Weber09} & Hodgkin & $D_{95\%}$, $D_{107\%}$
 Target  & mean DVH  \\
    & lymphoma (10) & $D_{1\%}$, $D_{33\%}$, $D_{50\%}$ & $D_{1\%}$,
$D_{99\%}$, $V_{95\%}$, $V_{107\%}$, $D_{mean}$  PTV \\
    &               & also give priority for each & $D_{integral}$
OAR   \\
   & & &  $D_{n\%}$, $V_{nGy}$ OAR   \\ \hline
 Bignardi {\it et al}\cite{Bignardi09} & abdominal & $D_{min} > 36 Gy$
 & $D_{99\%}$, $D_{1\%}$ PTV  \\
    & lymph nodes (14) & $D_{95\%}$, $D_{107\%}$ Target & $V_{95\%}$,
$V_{107\%}$ PTV  \\
    & & $D_{max}$ spine & HI, CI Target  \\
    & & $V_{15Gy}$ kidney  & EUD CTV and PTV  \\
    & & $V_{36Gy}$ duodenum, stomach, bowel & NTCP for OAR  \\
    & & $V_{15Gy}$ liver & $D_{1\%}$ I.P.O. objectives  \\
    & & $D_{0.5cm^3}$ duodenum, stomach bowel  &  \\
    & & last one is secondary importance &  \\ \hline
 Wu {\it et al}\cite{Wu09} & spine (10) & $V_{90}$ Target & $D_{x\%}$ PTV 
 \\
    & & $V_{xGy}$, $D_{max}$ depending & CI PTV  \\
    & & on nearby OAR & $V_{xGy}$, $D_{x\%}$ OAR  \\
    & &  & not the same as IPO  \\ 
    & &  & EUD OAR  \\ \hline
 Verbakel {\it et al}\cite{Verbakel09} & head \& neck (12) & $V_{<95\%}$,
$V_{>107\%}$ PTV & $V_{<95\%}$, $V_{>107\%}$ PTV \\
   & & $D_{max}$ cord, brainstem & $CI$ PTVs  \\
   & & $D_{mean}$ parotid & Stand. Dev. PTV  \\
   & & $V_{high}$ mouth, larynx & $D_{mean}$ parotid, larynx, mouth  \\
   & & $D_{max}$ other OAR &  \\ \hline
\end{tabular}
\caption{$^1$ Normal tissue objectives, $^2$ OAR: organ at risk, $^3$
  HI: homogeneity index,   $^4$ CI: conformity index} 
\label{tab:summaries}
\end{center}
\end{sidewaystable}

\begin{table}
\begin{center}
\begin{tabular}{|l|l|} \\ \hline
Cited Paper & Methods of comparison \\ \hline
Ref.~\cite{Tang10}  &   DO values presented for each case \\
  &  DVH's for individual cases \\ \hline
Ref.~\cite{Yoo10} 
  &   two sided Student's t test of each DO \\
  &  DO means and averaged DVHs \\ \hline
Ref.~\cite{Popescu10} 
  &  two-sided paired t test for each DO \\
  &  DO means and ranges and averaged DVH \\ \hline
Ref.~\cite{Weber09} 
  &  Wilcoxon matched-paired signed-rank test of each DO \\
  &  DO  means and standard deviations and averaged DVHs \\ \hline
Ref.~\cite{Bignardi09} 
  &  paired, two-tailed Student's t test of each DO \\
  &  DO  means and standard deviations and averaged DVHs \\ \hline
Ref.~\cite{Wu09} 
  &  Wilcoxon signed rank test for each DO \\
  &  DO means and standard deviations and averaged DVHs \\ \hline
Ref.~\cite{Verbakel09} 
  &  Wilcoxon matched-pair signed rank test for each DO \\
  &  DO means \\ \hline
\end{tabular}
\end{center}
\caption{Methods of comparison using decision objectives (DO).}
\label{tab:methods}
\end{table}

\end{document}